\documentclass[twocolumn,twoside,slac_two]{revtex4}
\usepackage{graphicx}
\usepackage{fancyhdr}
\usepackage{graphics}
\usepackage{epstopdf}
\usepackage{textpos}
\begin{document}

%%%%%%%%%%%%%%%%%%%%%% WRITE THE TITLE HERE %%%%%%%%%%%%%%%%%%%
\title{\centering The ATLAS $b$-Jet Trigger}
%%%%%%%%%%%%%%%%%%%%%% WRITE THE AUTHOR HERE %%%%%%%%%%%%%%%%%
\author{
\centering
\begin{center}
P. Hansson Adrian, on behalf of the ATLAS Collaboration
\end{center}}
\affiliation{\centering SLAC National Accelerator Laboratory, 2575 Sand Hill Road, Menlo Park, CA 94025, USA}
%%%%%%%%%%%%%%%%%%%%%% WRITE THE ABSTRACT HERE %%%%%%%%%%%%%%%%
\begin{abstract}
The online event selection is crucial to reject most of the events containing 
uninteresting background collisions while preserving as much as possible the 
interesting physical signals. The $b$-jet selection is part of the trigger 
strategy of the ATLAS experiment and a set of dedicated triggers was 
contributing to the event selection for the 2011 running. The $b$-jets acceptance 
is increased and the background reduced by lowering jet transverse energy 
thresholds at the first trigger level and applying $b$-tagging techniques at 
the subsequent levels. Different physics channels, especially topologies 
containing more than one $b$-jet where higher rejection factors are achieved, 
benefit from using the $b$-jet trigger. An overview of the 
$b$-jet trigger menu and performance on data is 
presented.
\end{abstract}

\maketitle
\thispagestyle{fancy}


\section{Introduction}
The start of proton-proton collisions at the Large Hadron Collider 
(LHC) in 2010 opened up a new era of exploration at the high energy 
frontier. With proton-proton collisions at $\sqrt{s}=7$~TeV and 
instantaneous luminosities higher than $10^{33}$cm$^{-2}$s$^{-1}$ 
the LHC is a real discovery machine, in particular extending 
sensitivity for new physics into mass scales above a TeV and 
set to dominate the high energy physics scene for years to 
come. During 2011 the LHC definitely entered real 
physics running with almost routine delivery of integrated 
luminosities larger than 50~pb$^{-1}$ per day and yearly total 
of more than 5~fb$^{-1}$. 

Both in the Standard Model and many of its possible extension,
such as Supersymmetry~\cite{susy1}-\cite{susy7}, the third 
generation of quarks, i.e. the top and bottom ($b$) quarks, 
play an important role due to their relatively large mass. 
The $b$-quark is also important 
as one of the main decay channels for the proposed 
Higgs boson and top quark production is one of the major backgrounds 
to many new physics physics searches.

\section{The ATLAS $b$-Jet Trigger}
The identification of jets originating from $b$-quarks ($b$-jets) 
is a central piece in 
the rich physics program of the ATLAS detector at the LHC. The 
large (approximately 40~m long and 25~m in diameter) multi-purpose 
ATLAS detector~\cite{atlas_detector} 
has a hermetic design with a large muon spectrometer surrounding 
electromagnetic- and hadronic calorimeters and an inner 
detector using three 
different technologies to track charged 
particles up to $|\eta|=2.5$. Crucial for $b$-jet 
identification, the innermost pixel detector provides 
three space point measurements with 50$\times$400~$\mu$m pixels, 
where the innermost layer is located only 5~cm from the beam. 
Tracking is 
extended through the large area silicon micro-strip detector with 
four double-layered sensors (80~$\mu$m pitch) and the straw 
tube transition radiation tracker up to a radius of about 111~cm.

\begin{figure}
\includegraphics[width=65mm]{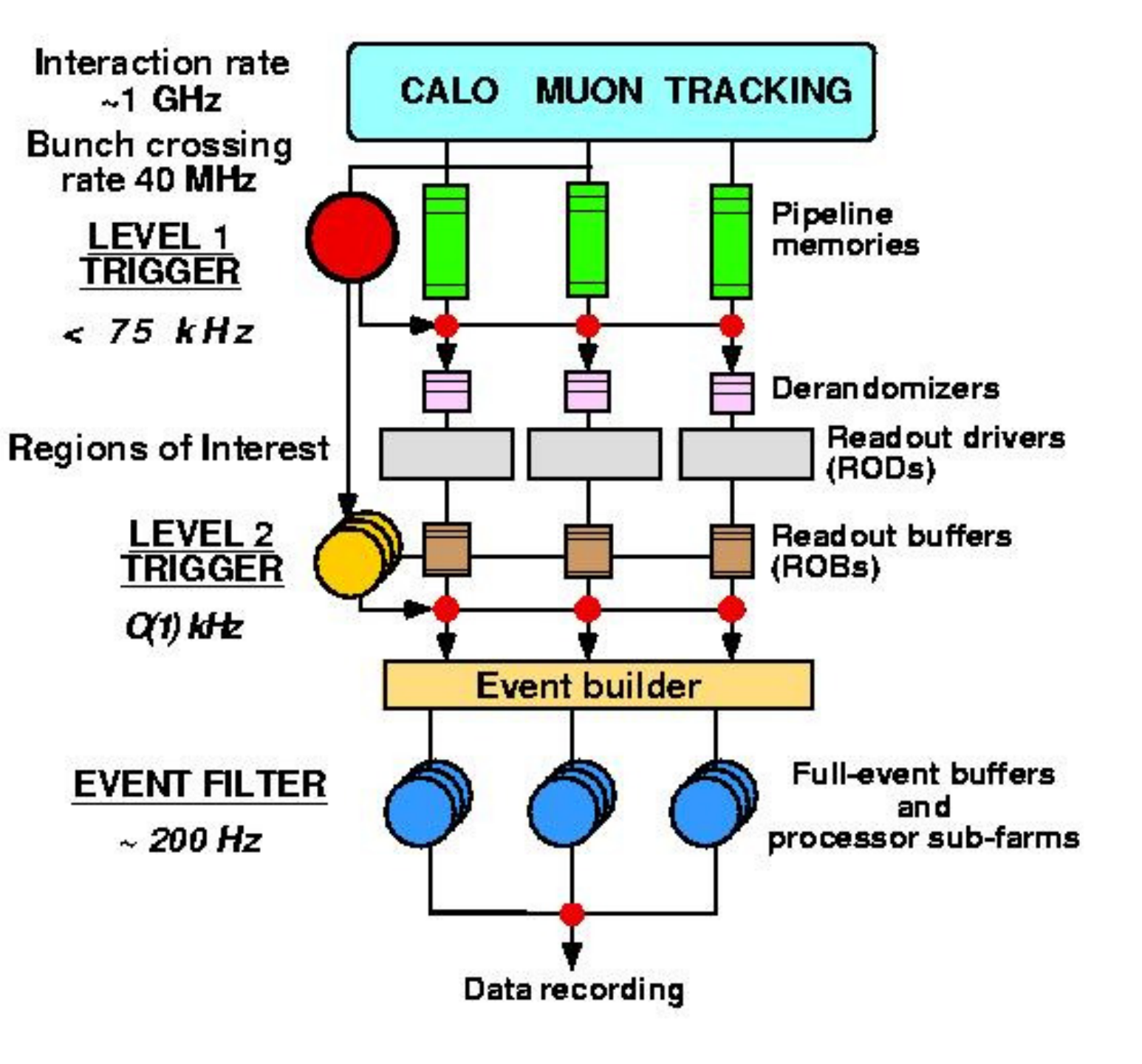}
\caption{Schematic overview of the trigger and data acquisition in ATLAS.}
\label{fig:tdaq}
\end{figure}

A challenge for the LHC experiments is the large interaction 
rate and multiplicities. The rate of events collected for further 
offline analysis needs to be reduced 
from the initial proton-proton bunch collision rate of 40~MHz, with more 
than 10 interactions per bunch-crossing, to $<400$~Hz while keeping 
the interesting events for further study. The ATLAS trigger 
system~\cite{tdaq} is a 3-tiered structure built 
around an initial custom made hardware trigger, Level 1 (LVL1), 
and two software-based levels collectively called the 
High Level Trigger (HLT), see Fig.~\ref{fig:tdaq}. 
The LVL1 trigger  builds trigger decisions using 
fast analog information from hits in the three-layered muon spectrometer 
and energy deposits in the calorimeter cells to identify signatures 
of high-$p_{\mathrm{T}}$ muons, electrons/photons, 
$\tau$-leptons and jets.
The ATLAS trigger system relies to a large extent on the so called 
Regions of Interest (ROIs) which are regions in the detector identified in 
the LVL1 trigger associated to a specific type of signature. These ROIs 
later form the basis for a more detailed reconstruction at the HLT, effectively 
restricting the amount of data needed to be shipped from the 
detector readout buffers. Since the $b$-jet trigger 
relies on information from the inner detector, 
the identification of $b$-jets can only start at the HLT. This makes 
the LVL1 jet reconstruction particularly important and an integral 
part of any $b$-tagging at the trigger level as it provides the seed ROI 
in which the inner detector tracking algorithms and subsequent 
$b$-tagging algorithms are executed. 
The LVL1 jet trigger is a fixed-size sliding 
window algorithm that sums energy in projective towers of size 
$\Delta\eta\times\Delta\Phi=0.4\times0.4$. The maximum event rate 
accepted at LVL1 is $\sim75$~kHz and the detector readout buffers need 
to receive the trigger decision no later than approximately 2.5$~\mu$s 
after the relevant bunch crossing. 

The HLT is a farm of mostly commercial computers and 
networking technology providing a fully configurable two-tiered 
trigger system seeded by the ROIs.
The Level 2 (LVL2) trigger accesses the full detector granularity 
and precision from the muon system, calorimeters and inner detector 
within the ROI and is optimized for speed to meet 
the maximum execution time of $\sim$ 40~ms. The HLT manages and 
steers the event during the LVL2 algorithm sequence execution 
and assigns data from accepted events to the final event building 
step. With an event building rate of about 3.5~kHz and a latency 
up to several seconds 
the Event Filter can run algorithms which are nearly 
identical to those used in offline event reconstruction and can 
limit the final output rate to below 400~Hz.

\subsection{The $b$-Jet Trigger Menu}
\begin{figure}
\includegraphics[width=65mm]{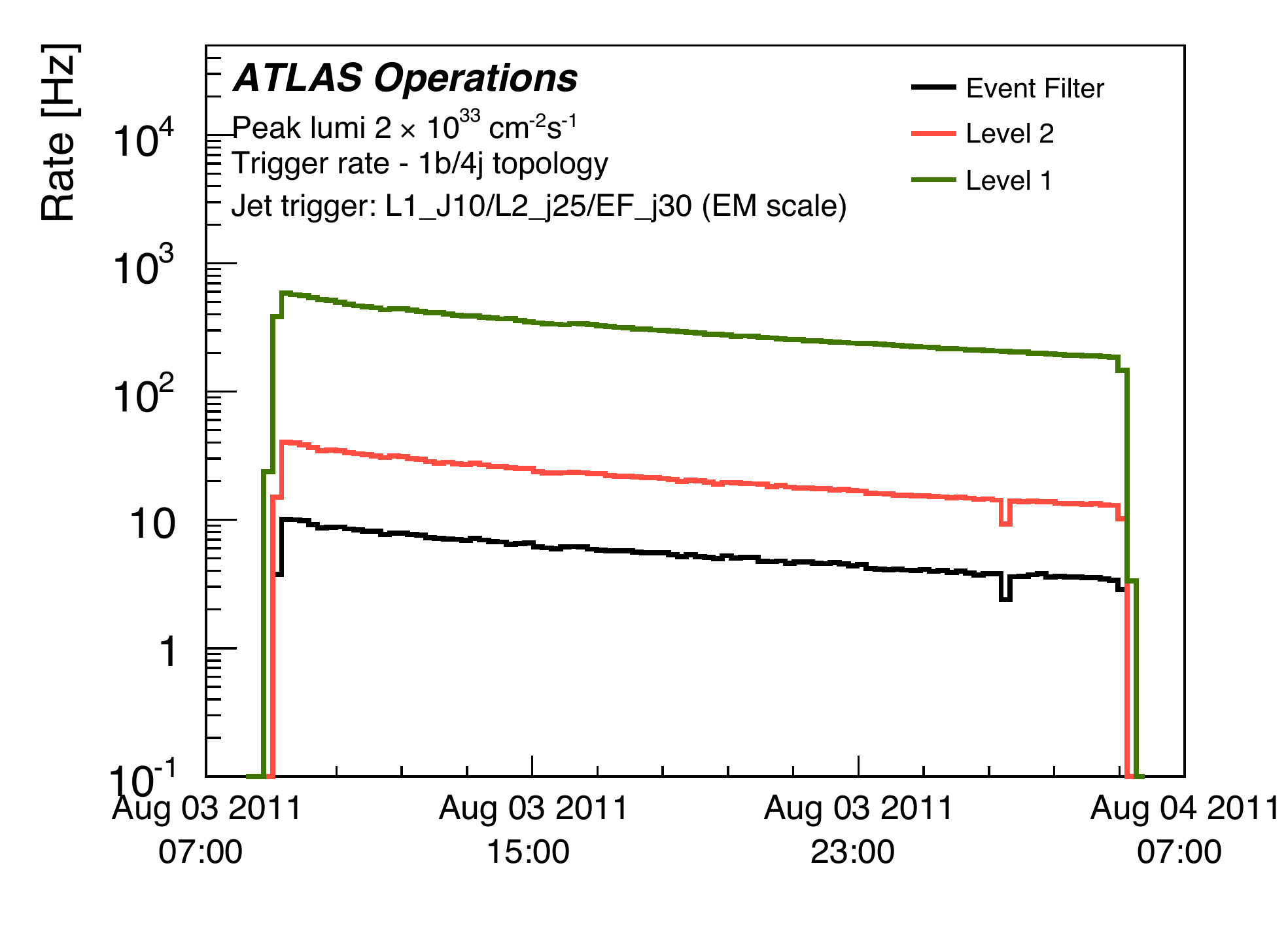}
\caption{Trigger rate at each trigger level during a typical LHC fill.}
\label{fig:rates}
\end{figure}

The $b$-jet trigger was actively selecting events during 2011. 
It consists not only of physics triggers to select 
signal events but also special 
triggers for monitoring and calibration purposes, including 
trigger efficiency measurements. 
The physics triggers were designed to cover a wide range of 
generic signals; a multi-jet trigger with one or 
more $b$-tags and a dijet trigger with two or more $b$-tagged 
jets. Typical rates at each trigger level can be seen in 
Fig.~\ref{fig:rates}~\cite{btrigwiki}.
%Tab.~\ref{tab:rates} and Fig.~\ref{fig:rates}.
%\begin{table}[t]
%\begin{center}
%\caption{Example of rates (in Hz) and rejection power of two type 
%of $b$-jet triggers for different signal topologies.}
%\begin{tabular}{|l|c|c|c|}
%\hline 
%\textbf{Trigger} & \textbf{L1}& \textbf{ LVL2}& \textbf{Event Filter} \\
%%&(Hz)&(Hz)&(Hz)\\
%\hline\hline
%$\geq1$ $b$-tags, $\geq4$ jets & 376 & 6 & 2  \\
%\hline
%$\geq2$ $b$-tags, $\geq2$ jets & 830 & 12 & 3  \\
%\hline
%\end{tabular}
%\label{tab:rates}
%\end{center}
%\end{table}

\subsection{The $b$-Tagging Algorithm}
\begin{figure}[htbp]
  \centering
\includegraphics[width=65mm]{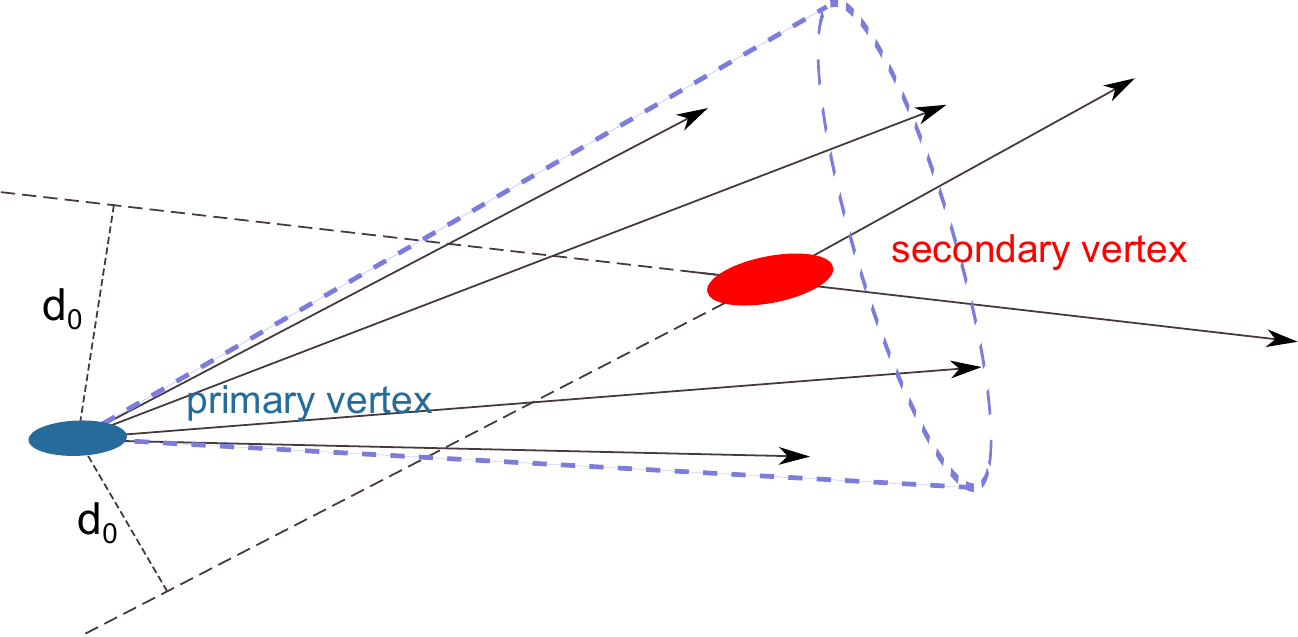}
  \caption{Schematic view of the tracks in a $b$-jet. }
\label{fig:bjet}
\end{figure}

The most natural choice in building a 
discriminating variable between $b$- and 
light jets\footnote{Light jets are defined here as jets originating 
from quarks from the two first generations or a gluon.}
is to exploit 
the transverse impact parameter, $\mathrm{d}_{0}$, defined as the 
distance of closest approach between the particle track and the 
primary vertex. 
The finite lifetime of $B$ hadrons ($\tau \approx 1.6$~ps, 
$c\tau \approx 490~\mu m$) allow them to 
typically travel a few mm before decaying. This, together with 
their relatively large mass produce tracks with on average 
large impact parameter compared to tracks originating from 
light jets (see Fig.~\ref{fig:bjet}). 
Normally also the measured uncertainty $\sigma$(d$_0$) is 
used to define the impact parameter significance, 
S(d$_0$)$=$d$_0/\sigma$(d$_0$),  
to better judge the likelihood of the track displacement.

\begin{figure}
\includegraphics[width=65mm]{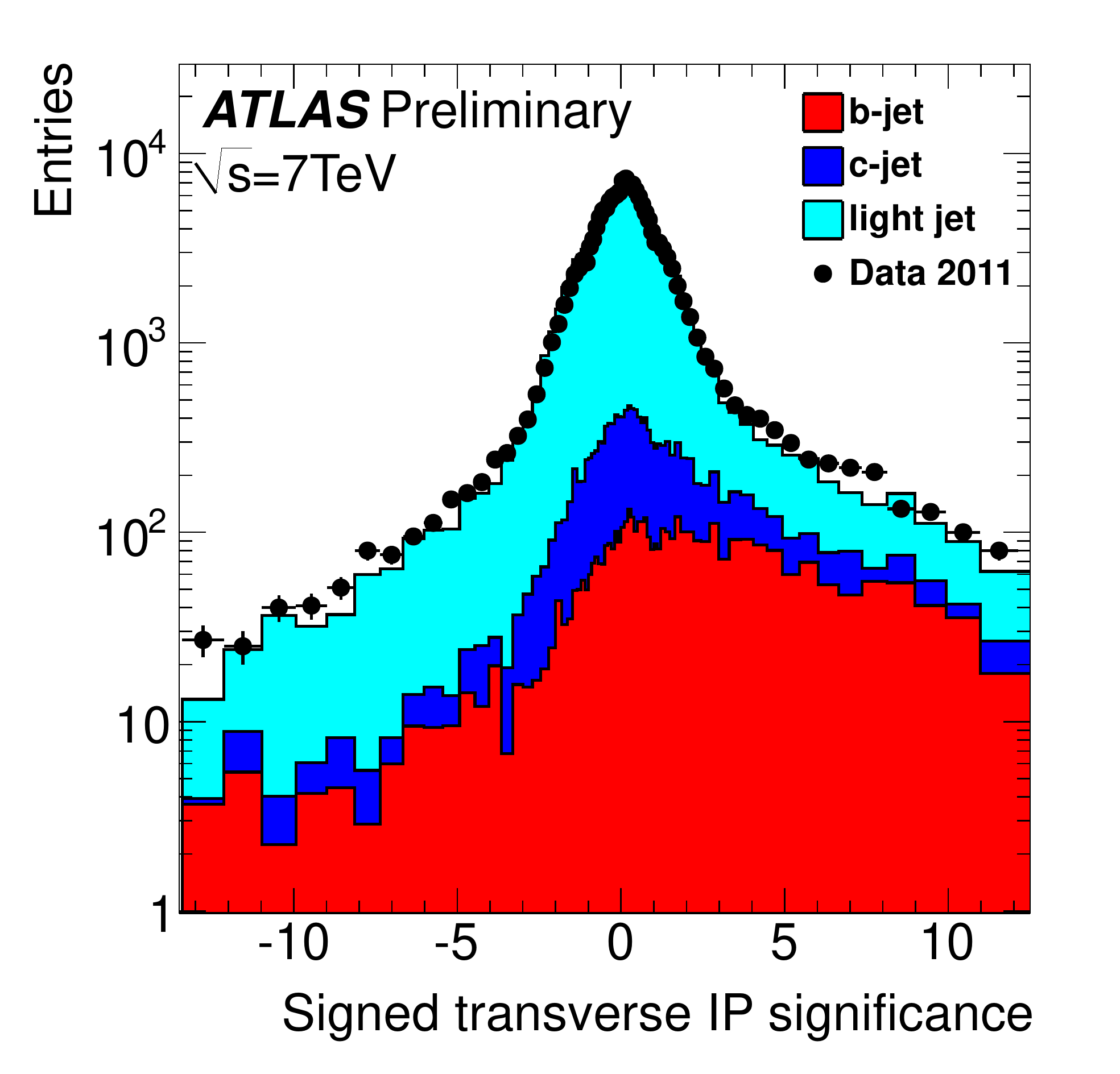}
\caption{The signed impact parameter comparing simulation and data.} 
\label{fig:ip}
\end{figure}

In 2011 the so-called {\it JetProb} algorithm was used 
to select $b$-jets in ATLAS at both LVL2 and Event Filter. 
This technique was first developed by the 
ALEPH collaboration~\cite{jetprob_ref} and 
then adopted in experiments at the Large Electron Positron 
Collider and the Fermilab Tevatron. 
The JetProb method computes the probability for a jet to 
originate from the primary 
vertex using the signed transverse impact parameter signiﬁcance 
of tracks associated with the jet. The sign of the impact 
parameter significance is determined from whether or not 
the track crosses the jet axis in front of the 
primary vertex (positive) or behind it (negative). 
Most of the tracks produced from decays of particles 
with long lifetime, such as a $B$ hadron, are positive.  
Due to finite impact parameter resolution, tracks in light 
jets, even if they originate from the primary vertex, 
may seem displaced but would have both signs roughly with equal 
probabilities. Figure~\ref{fig:ip} 
shows a comparison of the impact parameter significance for 
tracks associated 
to jets classified as $b$-, $c$- or light jets compared 
to that measured in data~\cite{btrigwiki}.
Each track is assigned a probability, $P$,
\begin{equation}
P = \int_{-\infty}^{-|\mathrm{d}_{0}/\sigma(\mathrm{d}_{0}|)}R,\label{eq:track_prob}
\end{equation}
where $R$ is the parameterization of the negative impact 
parameter resolution for tracks originating from the primary vertex. 
The resolution function can be determined from experimental data 
using the negative side of the signed impact parameter distribution, 
assuming the contribution from heavy-flavor particles is negligible. 
The individual track probabilities, $P_{i}$, are then combined 
into a per-jet quantity, $P_{jet}$,
\begin{equation}
P_{\mathrm{jet}} = P_{0} \sum_{\mathrm{i}=0}^{N_{\mathrm{tracks}}-1}\frac{(-\ln{P_{0}})^{\mathrm{i}}}{i!} \label{eq:jet_prob}
%\frac{ (-\log{P_{0})^{i}}{i!}, 
\end{equation}
where $P_{0}=\prod_{\mathrm{i}}P_{\mathrm{i}}$. 
With the assumption that no long-lived particles 
contribute to the selected tracks, $P_{\mathrm{jet}}$ has an expected 
uniform distribution between zero and one 
while tracks from jets with a long-lived particle decay 
tend to give $P_{\mathrm{jet}}$ closer to zero. 
This $b$-tagging algorithm is considered robust as it 
relies only on the knowledge of the 
negative transverse impact parameter distribution of 
prompt tracks in multi-jet events, rather easily derived 
from data. 

More sophisticated algorithms, such as explicit reconstruction 
of the secondary decay vertex, show promising improvements 
but are not yet used to actively select events at the 
trigger level.

\subsection{Primary Vertex and Beam Spot}
\label{sec:beamspot}
The primary vertex position is a vital ingredient 
for good $b$-tagging performance. A measurement of the primary vertex 
position depends strongly on the track multiplicity 
at the vertex. When the $b$-tagging algorithm is executed 
only tracks within the single ROI are available which degrades the 
primary vertex resolution in the transverse 
direction.
%\footnote{Dedicated studies on how to combine tracks 
%and vertices from multiple ROIs in the same event is ongoing.}. 
Note that in the longitudinal direction the resolution is less 
critical for the JetProb algoritm as the longitudinal impact 
parameter is only used to reject tracks 
from additional pile-up interactions.

\begin{figure}
\includegraphics[width=65mm]{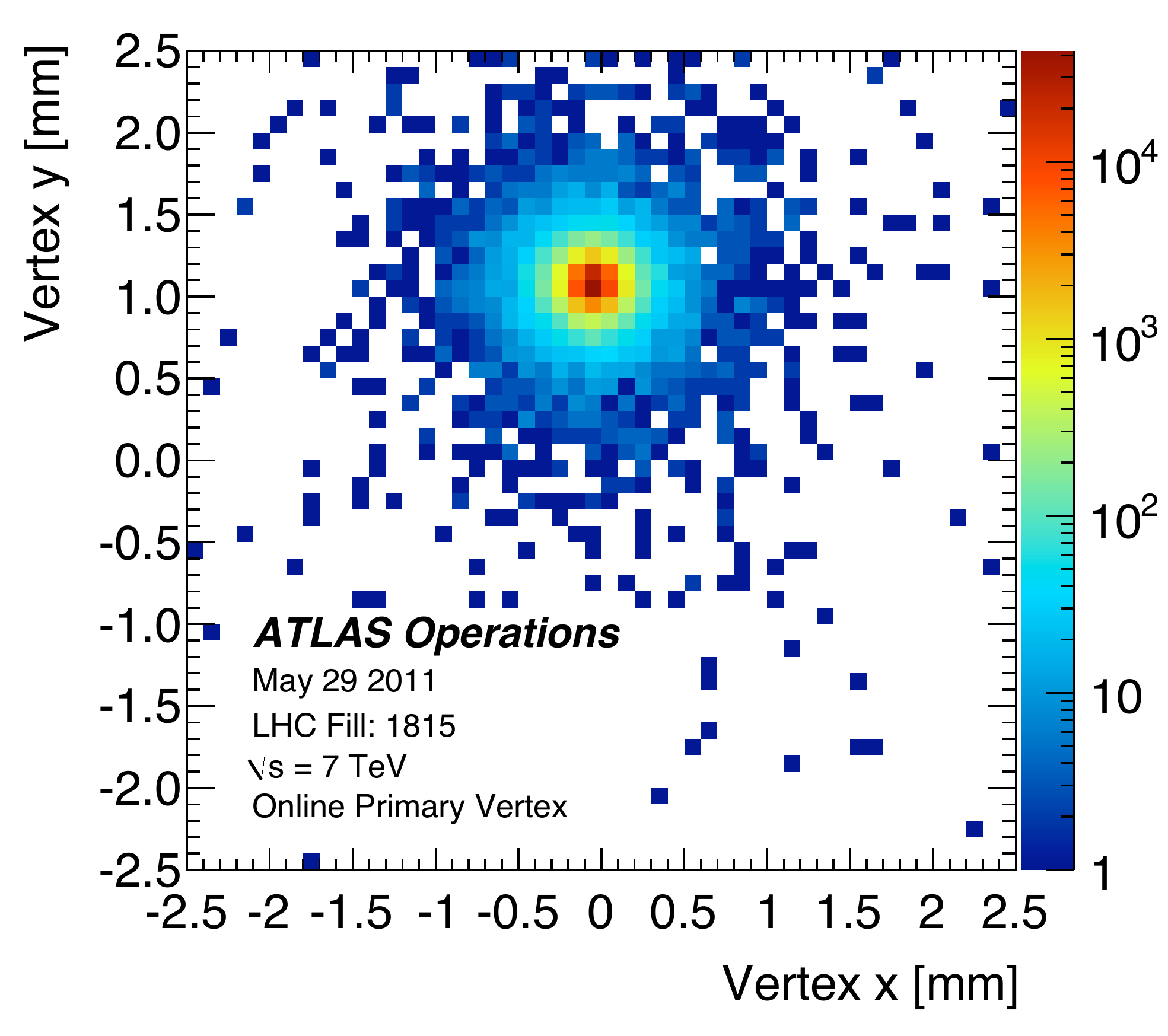} %Run182787_X_Resolution_and_Width.pdf}
\caption{The transverse distribution of primary vertices corresponding to 
about 1~min of data taking.}
\label{fig:BS_width}
\end{figure}

The solution is to exploit the online measurement of the relatively 
small beam spot every few minutes during data taking. 
Using special prescaled triggers, the per-event vertex position is 
integrated to measure the average position and shape 
(width and tilt) of the beam spot in intervals as short as a few minutes. 
%and made available to the other trigger algorithms running in the HLT.  
The width of the beam spot (typically around 25$~\mu$m at the
end of 2011) is estimated using a split vertex method~\cite{bs} 
online that exploits the measured distance between 
two fitted vertices, each recontructed with half of the tracks 
used in the original vertex fit. 
Figure~\ref{fig:BS_width} shows an example of the 
distribution of primary vertices used in the extraction 
of the beamspot parameters~\cite{bs}. 
The transverse beam spot position and tilt in the longitudinal direction 
is used to correct the track coordinates and the measured transverse width, 
$\sigma\mathrm{(BS)}$, to correct the track impact parameter 
uncertainty used in Eq.~\ref{eq:track_prob}: 
$\sigma'\mathrm{(d}_0\mathrm{)} = \sqrt{ \sigma\mathrm{(d}_{0}\mathrm{)}^{2} + \sigma\mathrm{(BS)}^{2} }$.
%\begin{equation}
%\label{eq:units}
%\sigma_{\mathrm{d}_0} = \frac{\rm track~\mathrm{d}_{0}}{ \sqrt{ \sigma_{\rm track~\mathrm{d}_{0}}^{2} + \sigma_{\mathrm{BS(x,y)}}}^{2}}.
%\end{equation}
The measured beam spot parameters are continiously monitored and 
updated whenever a significant change is detected.
%From the operational point of view the $b$-jet trigger algorithms 
%are enabled only after the beam spot measurement fulfills certain 
%quality requirements and it's beam spot position is updated whenever 
%a significant change is detected in either the position or width. 

\section{Bias On Offline $b$-Tagging}
\begin{figure}
\includegraphics[width=65mm]{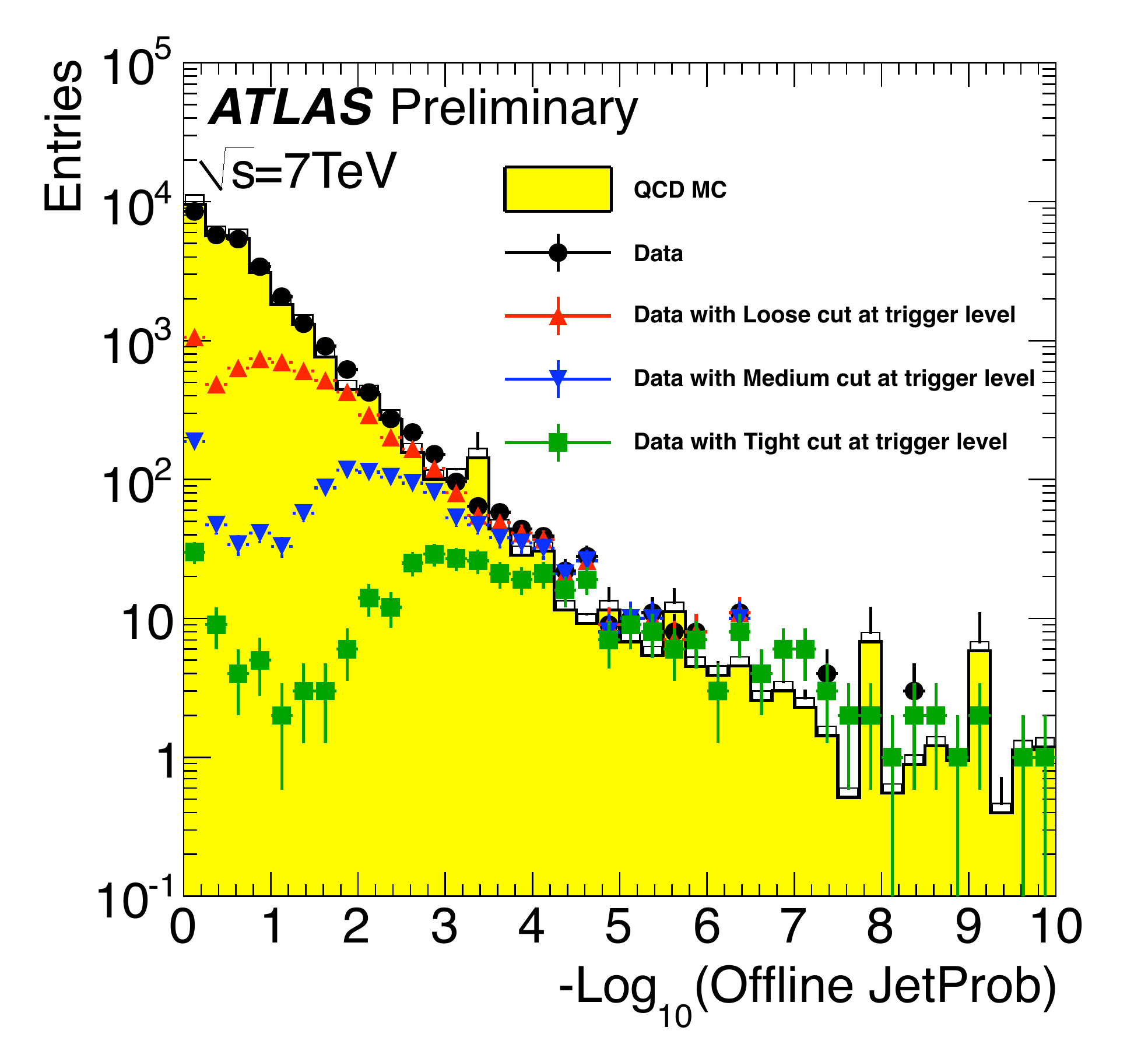}
\caption{The measured offline JetProb distribution for jets 
$b$-tagged at the HLT at three different operating points.}
\label{fig:jetprob_bias}
\end{figure}

One important aspect for physics analyses is to understand potential 
biases arising from trigger inefficiencies. For the $b$-jet trigger, 
the high rejection required together with 
the algorithm choice induces a non-negligible bias with respect to 
the offline $b$-tagging algorithms. 
Figure~\ref{fig:jetprob_bias} shows the bias on the 
offline JetProb weight distribution from three different 
operating points of the $b$-jet 
trigger corresponding to approximately 
90\%, 70\% and 50\% $b$-jet efficiency (w.r.t. jets tagged offline)~\cite{btrigwiki}. 
It's likely that the offline $b$-tagging algorithms will not 
be able to operate where the online $b$-jet tagging is 100\% 
efficient and still maintain a reasonably high combined 
online and offline $b$-jet efficiency. 
% efficiency is trigger efficiencyon the plateau. 
%In fact, for more sophisticated offline $b$-tagging algorithms 
%exploiting numerous variables to increase performance the correlation 
%between online and offline $b$-tagging will decrease further. 
Since detailed tracking information is notoriously hard to simulate 
this requires careful measurements of the combined online and offline 
$b$-tagging efficiency and mis-tag (light jets accidently tagged) 
rates in data. Such measurements are 
carried out using well-tested methods that explore 
muon properties from heavy flavor decays or explicit reconstruction 
of $B$ hadron decay chains.

\section{Conclusions and Outlook}
The $b$-jet trigger in ATLAS has been commissioned and 
actively rejecting events during 2011. 
Exploiting b-tagging information at the trigger level allows 
for the lowering of trigger thresholds for jets and missing energy. 
This leads to an increased selection efficiency for final states 
including $b$-jets, something which is used to improve the sensitivity 
in many physics analyses in progress.

To use the $b$-jet triggers together with an offline $b$-tagging 
requirement in physics analyses detailed measurements of the 
combined trigger and offline $b$-tagging efficiency and mistag 
rate are needed. This will properly take into account correlations 
between the trigger and offline requirements and correct for aspects 
which are not accurately described by the simulation of the 
trigger and the offline reconstruction. 

For 2012 many improvements are expected to be deployed to further 
improve the $b$-jet trigger performance. New $b$-tagging algorithms 
based on explicit reconstruction of secondary vertices will be used 
to improve light jet rejection. The improved jet measurements at 
the HLT compared with LVL1 will be used to refine the 
ROI direction and reduce the load on the data acquisition. Studies 
comparing new primary vertex algorithms at the trigger level may 
allow the usage of the an event-based primary vertex instead of 
the beamspot.

%\begin{acknowledgments}
%The authors wish to thank JACoW for their guidance in preparing
%this template.
%\end{acknowledgments}

\bigskip % extra skip inserted

\end{document}